\documentclass[]{article}
\usepackage{amsmath}
\usepackage{amssymb}
\usepackage{authblk}
\usepackage{array}
\usepackage{graphicx}
\usepackage{mathtools}
\usepackage{xcolor}

\title{\bf An iterative Monte Carlo method to solve nonlinear second-order differential equations}
\author[a]{Mart\'{\i}n Ch\'avez-P\'aez}
\author[a,*]{Enrique Gonz\'alez-Tovar}
\author[b]{Guillermo Iv\'an Guerrero-Garc\'{\i}a}
\affil[a]{Instituto de F\'{\i}sica, Universidad Aut\'onoma de San Luis Potos\'{\i}, \'Alvaro Obreg\'on 64, 78000 San Luis Potos\'{\i}, S.L.P., M\'exico}
\affil[b]{Facultad de Ciencias, Universidad Aut\'onoma de San Luis Potos\'{\i}, Av. Chapultepec 1570, Privadas del Pedregal, 78295 San Luis Potos\'{\i}, S.L.P., M\'exico}
\affil[*]{Corresponding Author: enrique.gonzalez@uaslp.mx}
\date{}

\begin{document}

\maketitle

\begin{abstract}
\noindent The Monte Carlo method is a thriving and mathematically beautiful numerical technique used extensively, nowadays, to deal with many demanding problems in diverse fields. Here, we present an iterative Monte Carlo algorithm to work out very general nonlinear second-order differential equations, with Dirichlet boundary conditions. An example of its usage is, also, reported.  
\end{abstract}

\section{Introduction}\label{introduction}
Since the conception of the Monte Carlo method (MCM) in its modern form \cite{Ulam1987}, this tantalising mathematical idea has become a powerful and successful route to solve extremely difficult and even ``intractable" \cite{Traub1994} problems in many areas of knowledge \cite{Hammersley1964}. In particular, the usage of the Monte Carlo probabilistic approach for the numerical solution of differential equations has a long history \cite{Courant1928} and represents one of its most useful applications in science and technology. In this regard, the amazing relation discovered between Brownian motion and potential theory \cite{Hersh1969} has prompted the employment of the random walk-based MCM to solve the very fundamental Poisson equation \cite{Hammersley1964,Sadiku2015}, i.e., $\nabla^2 \phi \left( \vec{r} \,\right) =-g$. Despite the importance of this subject, to the best of our knowledge, there exists no general Monte Carlo algorithm to solve the Poisson equation when $g$ also depends on $\phi \left( \vec{r} \,\right)$ and/or on its derivatives, apart from an explicit dependence on $\vec{r}$. In this context, below, we introduce and exemplify an {\it iterative} Monte Carlo method that allows to treat numerically ordinary differential equations of the form $\dfrac{d^2 y\left( x \right) }{dx^2}=F\left(  x, y\left( x \right), \dfrac{d y\left( x \right) }{dx}  \right)$, with Dirichlet boundary conditions. Notably, this reported probabilistic procedure can be readily expanded to deal with partial differential equations.   

\section{The method}\label{method}
A second-order ordinary differential equation (ODE) for $y(x)$, with Dirichlet boundary conditions, is generally stated as: 

\begin{equation}\label{ode2a}
	E\left(x, y\left( x \right), \dfrac{d y\left( x \right) }{dx}, \dfrac{d^2 y\left( x \right) }{dx^2}\right)=0,
\end{equation}  

\noindent such as $y(a)=y_a$ and $y(b)=y_b$, $ \ni\, a<b$.

\noindent If Eq. \eqref{ode2a} can be put in the form 

\begin{equation}\label{ode2b}
	\dfrac{d^2 y\left( x \right) }{dx^2}=F\left(  x, y\left( x \right), \dfrac{d y\left( x \right) }{dx}  \right), 
\end{equation}

\noindent the Iterative Monte Carlo Method (IMCM), discussed in the following, can be utilised to solve Eq. \eqref{ode2b}, for $y(a)=y_a$ and $y(b)=y_b$. As a preliminary, let us, then, consider a simpler case of such equation, i.e., 

\begin{equation}\label{ode3a}
	\dfrac{d^2 y\left( x \right) }{dx^2}=F\left( x \right) 
\end{equation}

\noindent (notice that the RHS term of the previous equation only contains an explicit dependence on the variable $x$). The usual random walk-based Monte Carlo algorithm to solve Eq. \eqref{ode3a} comprises the next steps \cite{Sadiku2015}:  

\medskip

\noindent STEP 1: Given a uniform mesh $\left\lbrace  x_0=a, x_1, x_2,..., x_{N-1}, x_N=b \right\rbrace $, for each free point $x_i$, such that $i=1,...,N-1$, generate a large number, $K$, of random walks, which start at $x_i$ and end when hitting an absorbing boundary site, either $x_0$ or $x_N$.

\smallskip

\noindent STEP 2: If the $j$-th walk arrives at the boundary after $m^{(i,j)}$ steps and has visited the sequence of locations $\left\lbrace  P_0^{(i,j)}=x_i, P_1^{(i,j)},P_2^{(i,j)},..., P_{m^{(i,j)}}^{(i,j)}=x_{end}^{(i,j)} \right\rbrace $, calculate the Monte Carlo estimator for $y(x_i)$ from 

\begin{equation}\label{estimator1}
	y(x_i)=\dfrac{1}{K}\sum_{j=1}^{K}  y(x_{end}^{(i,j)})-\dfrac{h^2}{2K}\sum_{j=1}^{K} \left\lbrace \sum_{l=0}^{\mu^{(i,j)}} F( P_l^{(i,j)})  \right\rbrace ,  
\end{equation}
 
\noindent where $\mu^{(i,j)}=m^{(i,j)}-1$ and $y(x_{end}^{(i,j)})$ being the value, $y_a$ or $y_b$, of the function $y(x)$ at the particular absorbing boundary point, $x_0=a$ or $x_N=b$, reached by the $j$-th random walk initiated at $x_i$. Observe that the approximate values of $y(x)$ for all the points in the mesh can be obtained for any arbitrary sequence of the indices in the set ${1,...,N-1}$, and, besides, that the computer implementation of the procedure can be easily parallelised.   

\medskip

\noindent Now, regarding the more general ODE stated by Eq. \eqref{ode2b}, in theory, the typical random walk methodology reviewed above can not be applied to work out such ODE, since the RHS of Eq. \eqref{ode2b} depends, precisely, on the unknown functions $y\left( x \right)$ and $y'(x)=\dfrac{d y\left( x \right) }{dx}$. However, and given that, at the end, these two sought functions depend on $x$, {\it the first version of our Iterative Monte Carlo Method} ({\it IMCM}) {\it proposes the use of an initial guess function, $y^{[0]}(x)$, with a well-known and simple dependence on $x$ in the interval $[x_0, x_N]$ and that fulfills the associated boundary conditions $y^{[0]}(x_0)=y_0=y_a$ and $y^{[0]}(x_N)=y_N=y_b$; this trial function will allow us to evaluate the last term in the RHS of Eq. \eqref{estimator1} and, consequently, to set up the ensuing practicable estimator for the value of the next guess function $y^{[1]}(x)$ at $x_i$}:

\begin{equation}\label{estimator2}
\begin{multlined}
  	y^{[1]}(x_i)=\dfrac{1}{K}\sum_{j=1}^{K}  y(x_{end}^{(i,j)})\,- \\
  	\dfrac{h^2}{2K}\sum_{j=1}^{K} \left\lbrace \sum_{l=0}^{\mu^{(i,j)}} F\left( P_l^{(i,j)}, y^{[0]}(P_l^{(i,j)}), {y^{[0]}}'(P_l^{(i,j)})\right)  \right\rbrace ,
\end{multlined}
\end{equation}

\noindent for $i=1,...,N-1$, with ${y^{[0]}}'(x_i)$ being the numerical derivative of $y^{[0]}(x)$ at each node (which can be calculated through any finite difference formula). After determining the values of $y^{[1]}(x)$ for all the free nodes, the corresponding numerical derivative has to be got to proceed with successive approximations of $y(x)$, obtained via the iteration prescription 

\begin{equation}\label{estimator3}
	\begin{multlined}
		y^{[s+1]}(x_i)=\dfrac{1}{K}\sum_{j=1}^{K}  y(x_{end}^{(i,j)})\,- \\
		\dfrac{h^2}{2K}\sum_{j=1}^{K} \left\lbrace \sum_{l=0}^{\mu^{(i,j)}} F\left( P_l^{(i,j)}, y^{[s]}(P_l^{(i,j)}), {y^{[s]}}'(P_l^{(i,j)})\right)  \right\rbrace ,
	\end{multlined}
\end{equation}

\noindent complemented with a finite difference approximation of its derivative. In a similar way to the scheme associated to Eq. \eqref{estimator1}, the prior iteration expression can be used to get $y^{[s+1]}(x_i)$ for all the free points following any order of the mesh indices. Additionally, this first version of the IMCM, given by Eq. \eqref{estimator3}, can be likewise parallelised. 

\noindent On the other hand, based on the idea of successive displacement we can try now to improve our IMCM as follows: 

\medskip

\noindent STEP 1: Advance an initial guess function, $y^{[0]}(x)$, in $[x_0, x_N]$, that satisfies the boundary conditions of the problem and, then, initialise $\tilde{y}(x_i)=y^{[0]}(x_i)$, for $i=0,...,N$. 

\smallskip

\noindent STEP 2: For each free node, perform $K$ absorbing random walks starting at $x_i$ to obtain and update immediately $\tilde{y}(x_i)$ employing the formula 

\begin{equation}\label{estimator4}
	\begin{multlined}
		\tilde{y}(x_i)=\dfrac{1}{K}\sum_{j=1}^{K}  y(x_{end}^{(i,j)})\,- \\
		\dfrac{h^2}{2K}\sum_{j=1}^{K} \left\lbrace \sum_{l=0}^{\mu^{(i,j)}} F\left( P_l^{(i,j)}, \tilde{y}(P_l^{(i,j)}), \tilde{y}'(P_l^{(i,j)})\right)  \right\rbrace .
	\end{multlined}
\end{equation}

\smallskip

\noindent STEP 3: Right after calculating $\tilde{y}(x_i)$ in STEP 2, differentiate it numerically to produce $\tilde{y}'(x_i)$.

\smallskip

\noindent STEP 4: Repeat entirely the previous steps 2 and 3 to complete a cycle of $T$ iterations. 

\medskip

\begin{figure} [h]
	\includegraphics[width=\linewidth]{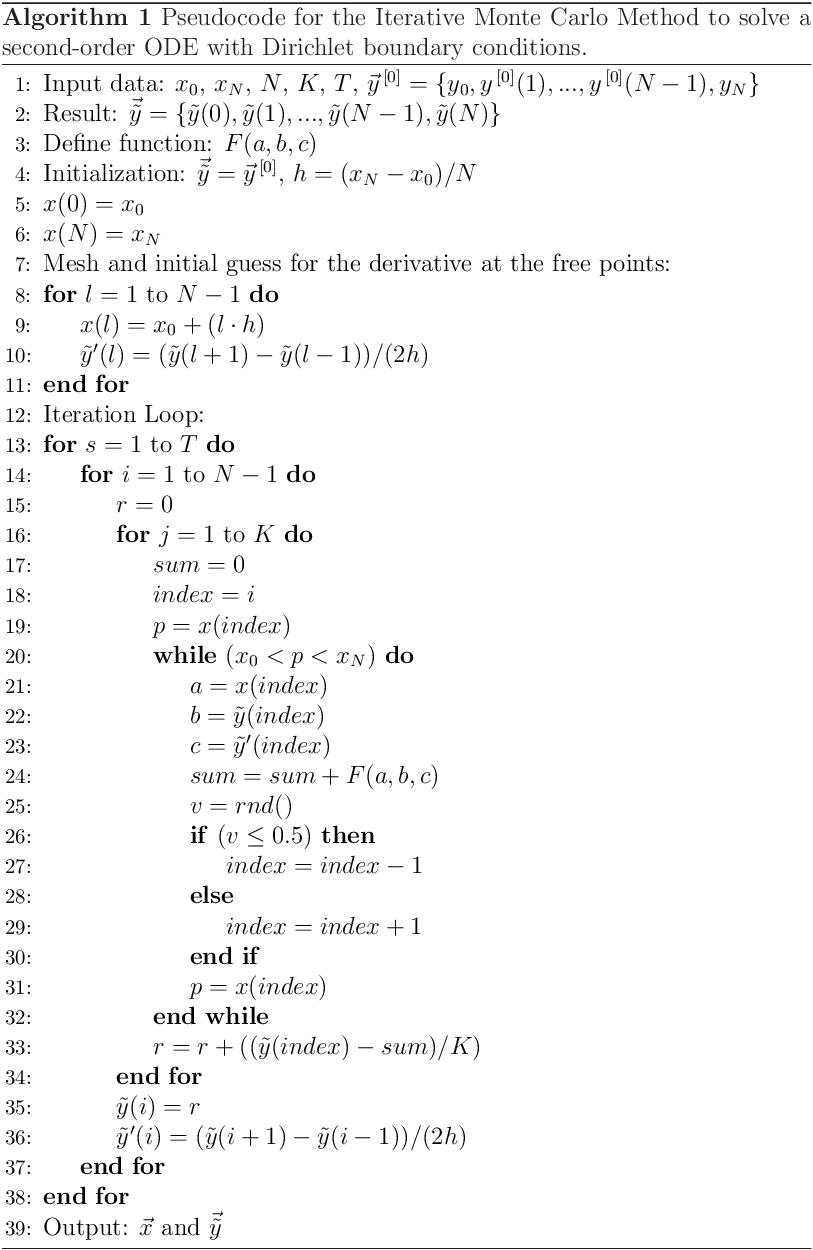}
	\caption{Pseudocode for the Iterative Monte Carlo Method (successive displacement version).}
	\label{pseudocode}
\end{figure}

\noindent This last stochastic and successive displacement process represents a refined, and seemingly hastened, version of the IMCM introduced here (the pseudocode of the corresponding algorithm can be consulted in Fig. \ref{pseudocode}).   

\section {An example of application}\label{example}
Let us consider the following nonlinear second-order ODE of the type $\dfrac{d^2 y\left( x \right) }{dx^2}=F\left(  x, y\left( x \right) \right)$ \cite{Yin2014}: 

\begin{equation}\label{ode01}
	\dfrac{d^2 y\left( x \right) }{dx^2}=\pi^2 \sin \left(\pi x \right) +e^{\sin \left(\pi x \right)+x}-e^{-y\left( x \right) },
\end{equation}

\noindent with $y\left( 0\right)=0$ and $y\left( 1 \right) =-1$. Eq. \eqref{ode01} has the exact solution 

\begin{equation}\label{exactsol}
	y^{exact}\left( x \right) = -\sin \left(\pi x \right)-x,
\end{equation}

\noindent which it will be utilised below to asses the performance of our approximate numerical IMCM results. 

\begin{figure} [h]
	\includegraphics[width=\linewidth]{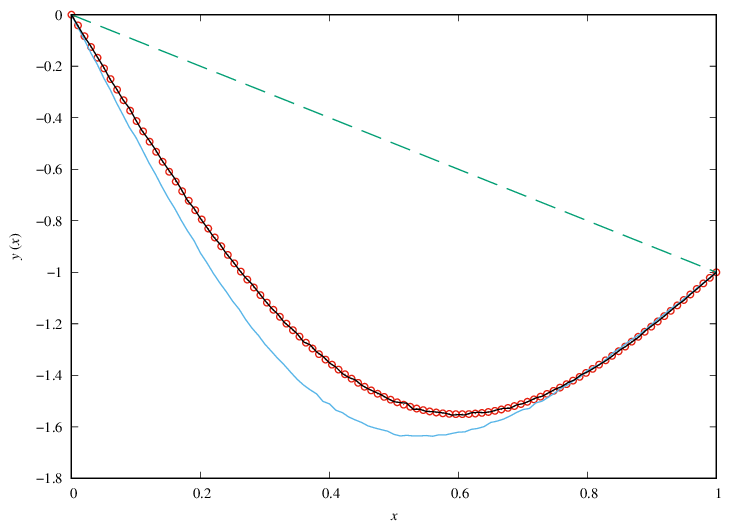}
	\caption{Solution to the ODE specified in Eq. \eqref{ode01}: exact solution given by Eq. \eqref{exactsol} (red open symbols); linear initial guess (green interrupted line); IMCM numerical approximate results after one iteration with Eq. \eqref{estimator4} (blue continuous line); IMCM numerical approximate results after ten iterations with Eq. \eqref{estimator4} (black continuous line).}
	\label{figure_m}
\end{figure}

\noindent Coding the enhanced IMCM procedure described at the end of Section \ref{method} (see Fig. \ref{pseudocode}), such that  

\begin{equation}\label{source01}
	F\left( x, y\left( x \right) \right) = \pi^2 \sin \left(\pi x \right) +e^{\sin \left(\pi x \right)+x}-e^{-y\left( x \right) },
\end{equation}

\noindent we obtained the data included in Fig. \ref{figure_m}. Therein, we show the exact solution given by Eq. \eqref{exactsol} (red open symbols), the linear initial guess (green interrupted line), the IMCM results after a first iteration with Eq. \eqref{estimator4} (blue continuous line), and the IMCM prediction after ten iterations with Eq. \eqref{estimator4} (black continuous line). In each Monte Carlo iteration we have employed a fixed grid in $\left[ 0,1 \right]$, with $N=100$ equally-sized partitions, combined with $K=10^5$ random walks, to estimate the value of $y(x)$ at each non-boundary node of the grid via Eq. \eqref{estimator4}. The error in the $s$-th iteration, $E^{\left( s \right)}$ is calculated from 

\begin{equation}\label{errork}
	E^{\left( s \right)}= \sqrt{\sum_{i=0}^{N} \left[ y^{exact} \left( x_i \right) - {\tilde y}^{\left( s \right) } \left( x_i \right) \right] ^2} \,,
\end{equation} 

\noindent with ${\tilde y}^{\left( s \right)} \left( x \right)$ being the $s$-th IMCM approximation. The corresponding errors for the IMCM results for iterations 1 and 10, plotted in Fig. \ref{figure_m}, are $1.001\times 10^0$ and $2.787\times 10^{-2}$, respectively.

\section{Conclusions}\label{conclusions} 
In this paper, we have proposed an Iterative Monte Carlo Method to solve second-order ODEs of the general form $\dfrac{d^2 y\left( x \right) }{dx^2}=F\left(  x, y\left( x \right), \dfrac{d y\left( x \right) }{dx}  \right)$, with Dirichlet boundary conditions. Our algorithm is based on the classic random walk approach, but it is enriched with an iterative process starting from a trial or guess function, which allows the evaluation of the RHS of the ODE. To illustrate its performance, we have also included an example of application. The presented IMCM can be extended straightforwardly to second-order partial differential equations.

\end{document}